\documentclass[tradiabstract]{aa}
\usepackage{xspace}
\usepackage{hyperref}
\usepackage{graphicx}
\usepackage{txfonts}
\newcommand{\xmm}{\textit{XMM-Newton}\xspace}
\newcommand{\swift}{\textit{Swift}\xspace}
\graphicspath{{figures/}}

\begin{document}

   \title{Monitoring clumpy wind accretion in supergiant fast X-ray transients with XMM-Newton}

   \author{C. Ferrigno
          \inst{1}
          \and
          E. Bozzo\inst{1}
          \and
          P. Romano\inst{2}
          }

   \institute{Department of astronomy, University of Geneva, chemin d'\'Ecogia, 16, CH-1290, Versoix, Switzerland\\
              \email{carlo.ferrigno@unige.ch}
              \and
              INAF, Osservatorio Astronomico di Brera, Via E. Bianchi 46, I-23807, Merate, Italy
             }

   \date{}

  \abstract
   {Supergiant fast X-ray transients (SFXTs) are a sub-class of supergiant high mass X-ray binaries hosting a neutron star accreting from the stellar wind of a massive OB companion. Compared to the classical systems, SFXTs display a  pronounced variability in X-rays that has long been (at least partly) ascribed to the presence of clumps in the stellar wind. We report here on the first set of results of an on-going XMM-Newton observational program aimed at searching for spectroscopic variability during the X-ray flares and outbursts of the SFXTs. The goal of the paper is to present the observational program and show that the obtained results are according to expectations, with a number of flares (between one and four) generally observed per source and per observation (20~ks-long, on average). We base our work on a systematic and uniform analysis method optimized to consistently search for spectral signatures of a variable absorption column density, as well as other parameters of the spectral continuum. 
   Our preliminary results
   show that the program is successful and the outcomes of the analysis support previous findings that most of the X-ray flares seem associated to the presence of a massive structure approaching and getting accreted by the compact object.  However, we cannot rule out that other mechanisms are  at work together with clumps to enhance the X-ray variability of SFXTs. This is expected according to current theoretical models. The success of these observations shows that our observational program can be a powerful instrument to deepen our understanding of the X-ray variability in SFXTs. Further observations will help us in achieving a statistically robust sample. This is required to conduct, in the future, a systematic analysis on the whole SFXT class with the ultimate goal of disentangling the role of different mechanisms giving  rise to these events.}


   \maketitle
%

\section{Introduction}
\label{sec:intro}

Supergiant fast X-ray transients (SFXTs) are a sub-class of the so-called classical supergiant X-ray binaries (SgXBs) in which a neutron star (NS) accretes from the wind of its supergiant OB companion\footnote{So far, the presence of a neutron star instead of a black-hole in most of the SFXTs is based on the spectral energy distribution in the
X-ray domain that in all cases is remarkably similar to that of X-ray pulsars. Pulsations have been detected only in two out of the roughly ten known SFXTs.}. At odds with classical systems, SFXTs display a much more prominent X-ray variability, comprising sporadic hour-long outbursts reaching $\sim$10$^{37}$ erg s$^{-1}$. In between the outbursts, SFXTs spend long intervals of time at a lower luminosity state ($\sim$10$^{33-34}$~erg\,s$^{-1}$), during which very frequent fainter X-ray flares take place. These have a duration similar to that of the brightest outbursts and are characterized by similar spectral variations \citep[see, e.g.,][for a recent review]{nunez17}.

Extrapolating from the widely accepted interpretation that the X-ray variability of classical SgXBs is mostly associated with accretion from the clumpy wind of the OB
supergiant \citep[see, e.g.,][and references therein]{walter2015}, it was originally proposed that that the SFXT phenomenology could be explained by invoking wide elongated
orbits and the presence of extremely dense clumps in their stellar winds \citep[see, e.g.,][]{zand05,negueruela10,walter07}. This idea has been challenged by the discovery of SFXTs with short orbital periods (3--5~days) and the lack of any evidence supporting a dichotomy in the properties of stellar winds in classical SgXBs and the SFXTs \citep{lutovinov13,romano14b,bozzo15,nunez17,pragati18,hainich}.

The currently most agreed scenario to explain the SFXT behavior is the simultaneous combination of mechanisms halting the accretion onto the compact objects for most of the time \citep[through, e.g., magnetic/centrifugal gating or assuming a long-lasting ``subsonic settling accretion regime'';][]{bozzo08,shakura12,shakura14} together with  moderately dense clumps which can temporarily restore accretion when they are captured by the gravitational field of the compact object  \citep[see also][]{mellah17, mellah18}.

The presence of clumps around the neutron stars in SFXTs can be observationally tested. A clump passing in front of the NS without being accreted is expected to (at least partly) obscure the X-ray source and its presence can thus be revealed by the signature of photoelectric absorption. Clumps that are instead (at least partly) accreted by the NS lead to a temporarily larger mass accretion rate, giving rise to X-ray flares or outbursts characterized by an enhanced local absorption \citep[see, e.g.,][and references therein, BZ17 hereafter]{Bozzo2017}. Events of source dimming associated with the passages of clumps have been reported in several observations of SFXTs \citep[see, e.g.,][]{rampy09,drave13}, while a systematic investigation of the signatures of clumps during flares and outbursts of SFXTs has been presented by BZ17 exploiting all \xmm\ observations available up to 2017.
\begin{table*}
	\caption{\label{tab:observations}Log of all observations used in this paper. We reported for completeness the total available exposure time for each observation after the application of the good time intervals, the estimated flux of time averaged spectra, as well as the energy $E_{\rm HR}$ used to separate the soft and high energy bands in the computation of the HR  (see Sect.~\ref{sec:data_analysis} for details).}
	\centering
	\begin{tabular}{llllccc}
		\hline
		\hline
		OBSID & Target & TSTART & TSTOP & Exp.\tablefootmark{a} & flux\tablefootmark{b} & $E_\mathrm{HR}$ \\
		& & UTC & UTC & ks & $10^{-12}$ cgs & keV\\
		\hline
0823990201 & IGR\,J18483$-$0311 & 2019-04-07 03:40:03 & 2019-04-07 13:33:11 & 23.3 & 47.9 & 3.0 \\
0823990301 & IGR\,J11215$-$5952 & 2018-12-30 12:46:57 & 2018-12-30 20:50:37 & 17.1 & $<$0.006 \tablefootmark{c} & --\\
0823990401 & IGR\,J16418$-$4532 & 2019-02-21 06:06:15 & 2019-02-21 12:37:19 & 16.3 & 41.9 & 3.5 \\
0823990501 & AX\,J1949.8+2534 & 2018-10-25 17:40:29 & 2018-10-26 01:47:21 & 22.3 & 0.8 & 3.72 \\
0823990601 & IGR\,J16479$-$4514 & 2018-08-28 06:19:06 & 2018-08-28 14:01:46 & 9.3 & 13.3 & 4 \\
0823990801 & IGR\,J18462$-$0223 & 2018-10-21 17:00:27 & 2018-10-22 02:01:25 & 15.3 & 0.2 & 3.5 \\
0823990901 & IGR\,J16328$-$4726 & 2018-09-18 15:26:40 & 2018-09-18 22:02:40 & 14.3 & 2.2 & 4 \\
0823991001 & SAX\,J1818.6$-$1703 & 2019-03-13 17:04:48 & 2019-03-13 23:24:12 & 18.9 & 3.0 & 4 \\
0844100101 & IGR\,J18410$-$0535 & 2019-10-18 17:52:09 & 2019-10-18 23:21:30 & 17.0 & $<$0.02 \tablefootmark{c} & --\\
0844100701 & IGR\,J18450$-$0435 & 2019-09-27 08:37:04 & 2019-09-27 16:49:43 & 24.0 & 2.7 & 3.5 \\
\hline
\hline
	\end{tabular}
	\tablefoot{
		\tablefoottext{a}{EPIC-pn exposure after filtering.}
		\tablefoottext{b}{Absorbed flux in the 1--10~keV energy band.}
		\tablefoottext{c}{We set an upper limit at 90\% confidence level on the unabsorbed flux, as the target is not detected in this observation. See Sect.~\ref{sec:upper-limits} for details on the computation.}
	}
\end{table*}

As shown by BZ17, the unique advantage of \xmm\  \citep{jansen01aa} is to provide the largest available effective area in the soft X-ray domain to search for variations in the spectral properties of the X-ray emission from SFXTs during the rise and decay from their flares/outbursts on time scales comparable to the dynamics of the clumpy wind accretion (i.e., $\sim$ few 100--1000\,s). For the analysis of all \xmm\  data, BZ17 adaptively rebinned the energy resolved lightcurves of all sources and used the measured hardness-ratio variations to drive the selection of different time intervals for the spectral extraction. The first interesting outcome of their study was the apparent lack of a correlation between the dynamic range in the X-ray flux and in the absorption column density achieved by any of the observed sources. This was interpreted as an important indication that accretion inhibition mechanisms are at work in the SFXTs and clumps cannot be the only ingredient to explain their extreme X-ray variability. The second outcome was a tentative evidence that lower absorption column densities are measured at higher fluxes compared to low/intermediate fluxes. If confirmed, this could prove that accreted clump gets photo-ionized at the peaks of flares/outbursts and thus the X-ray spectral properties during these states can be used as a probe of the ionization status and density of the clump.

In order to consolidate the previous findings by BZ17, we initiated in 2018 an observational program with \xmm\ to catch and perform HR-resolved spectral analyses of as many flares as possible from all known SFXTs. The goal is to collect over the coming years a statistically meaningful sample of flares, e.g., at least $\sim$25--30 for each source and be able in the future to carry out a full statistical investigation of the properties of the flares across the entire SFXT class (as indicated by BZ17). Long term studies of the SFXTs have shown that an X-ray flare of moderate intensity occurs on average every few ks (BZ17, \citealt{romano14}) and therefore 20~ks-long \xmm\ observations can be very effective in significantly increase the presently available database of these events observed with the required sensitivity and spectral resolution.

In this paper, we report on the first ten observations obtained between 2018 and 2019 from our on-going \xmm\ SFXT monitoring with the goal of illustrating the potentialities of the proposed program to be continued in the years to come until the full statistically meaningful sample of flare will be available for each of the SFXT  sources. We also show here that the availability of several observations performed in the same mode (e.g., with the same \xmm\ instrument set-up) allowed us to develop a largely automatized and standardized analysis, obtaining easily comparable results among the different flares from all observed sources. We present our newly developed data analysis process in Sect.~\ref{sec:data_analysis}, together with a summary of all observations considered for this paper. We then describe our results in Sect.~\ref{sec:results} and discuss them in Sect.~\ref{sec:discussion}. We do not include in this paper a summary of the literature results on the considered SFXT sources, as these were already presented in recent reviews \citep{nunez17} and in BZ17. This paper will serve as a reference for the future developments of our monitoring program. and to support the extension of the program in the years to come.

\section{Observations and data analysis}

All observations that have been considered for this paper are listed in Table~\ref{tab:observations}. We have a total of ten observations performed in the direction of as many confirmed SFXT sources. This is the ensemble of the observations obtained since 2018 out of our ongoing monitoring program. All observations have total exposure times of roughly 20~ks (before filtering for the high background intervals, see Sect.~\ref{sec:data_analysis}). We provide the details of the analysis of all observations in the following sub-sections.

\subsection{Data analysis process}
\label{sec:data_analysis}

For all observations used in this paper the EPIC pn \citep{Struder2001} was operated in full imaging mode, the  EPIC-MOS1  \citep{Turner2001} in small window, and the EPIC-MOS2 in timing mode. This instrument set-up was chosen to cope with the known unpredictably large changes in the X-ray luminosity of the SFXTs within the time scale of several ks. We extracted cleaned event lists in each observation using the \texttt{epproc} and \texttt{emproc} tools provided within the SAS software (version \texttt{xmmsas\_20190531\_1155}), respectively for the EPIC pn and EPIC-MOS cameras (we adopted default parameters for both tools). Up-to-date calibration files (November 2019) were obtained from the \xmm\ repository and used for the data processing and the following analysis. We applied a uniform data reduction and analysis recipes to all data sets. To maximize the scientific outcome of each single observation, we optimize some specific parameters as described in Sect.~\ref{sec:deviations}. This is related to well-known issues with the highly variable \xmm\ background, which inevitably had to be dealt with individually (although the analysis strategy for all observations remains the same). We verified that no usable RGS data were present for any of the observed sources due to the high absorption column densities that characterize their X-ray emission (see Sect.~\ref{sec:results}).
\begin{figure*}
    \centering
    \includegraphics[scale=0.45]{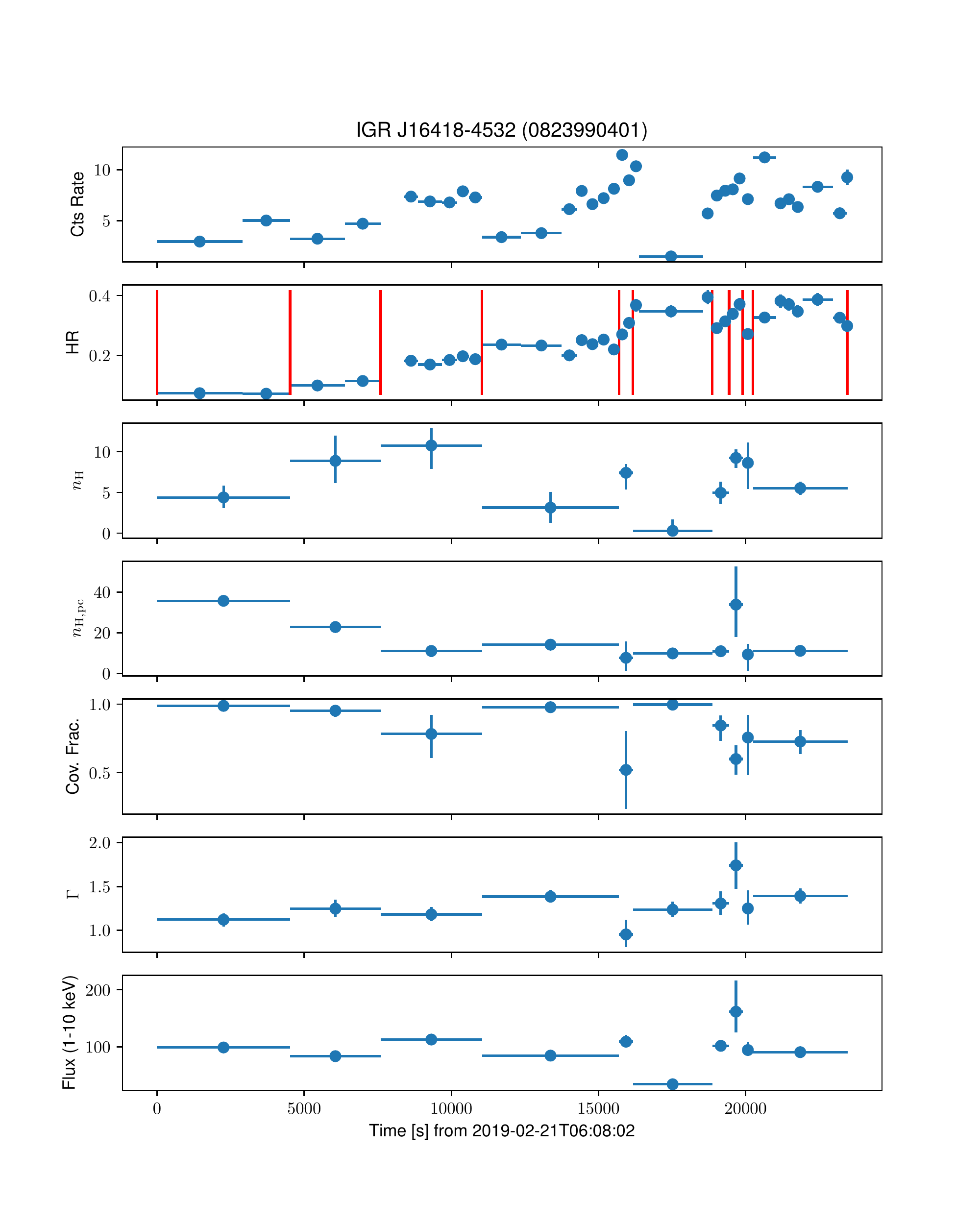}
    \includegraphics[scale=0.45]{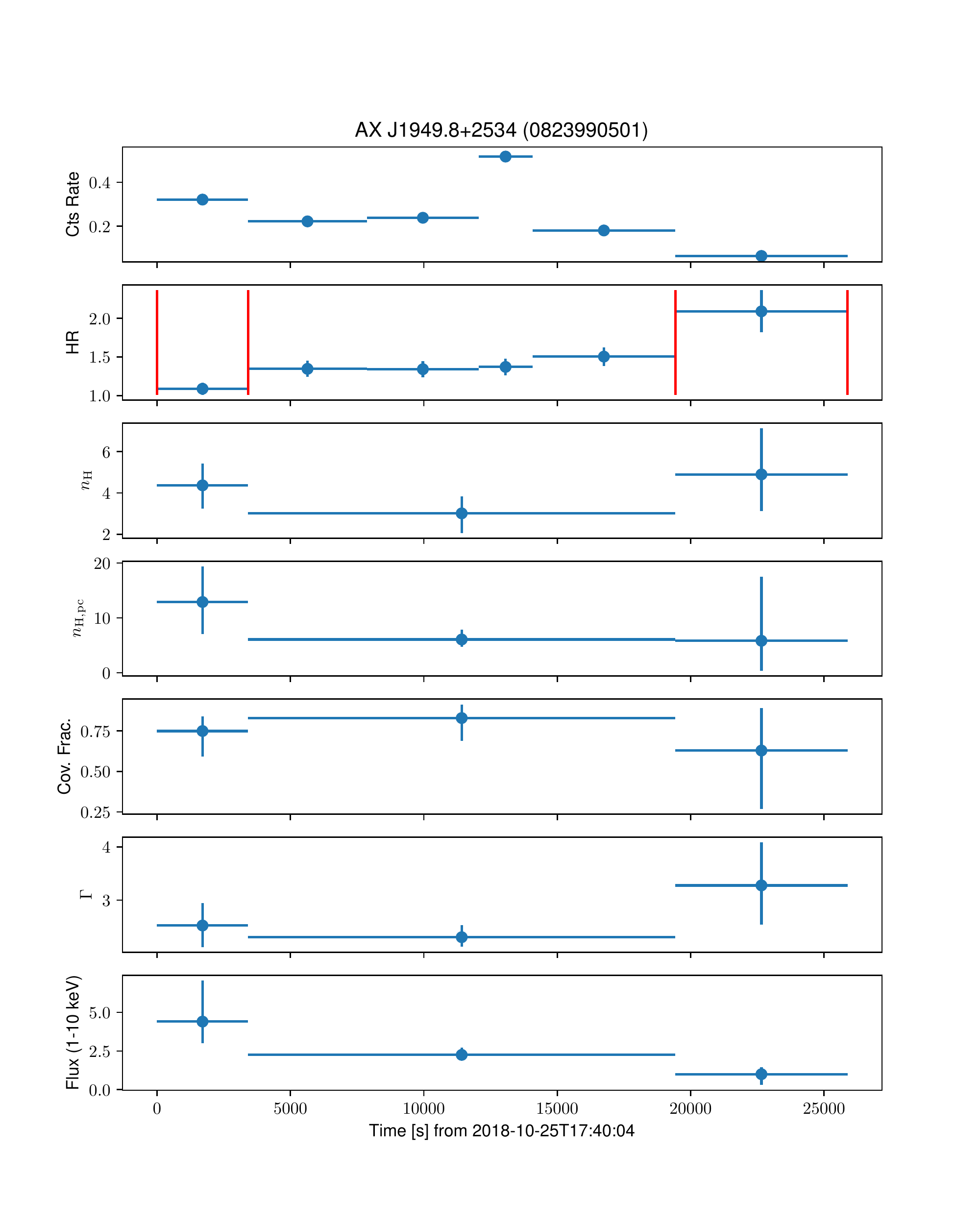}
    \includegraphics[scale=0.45]{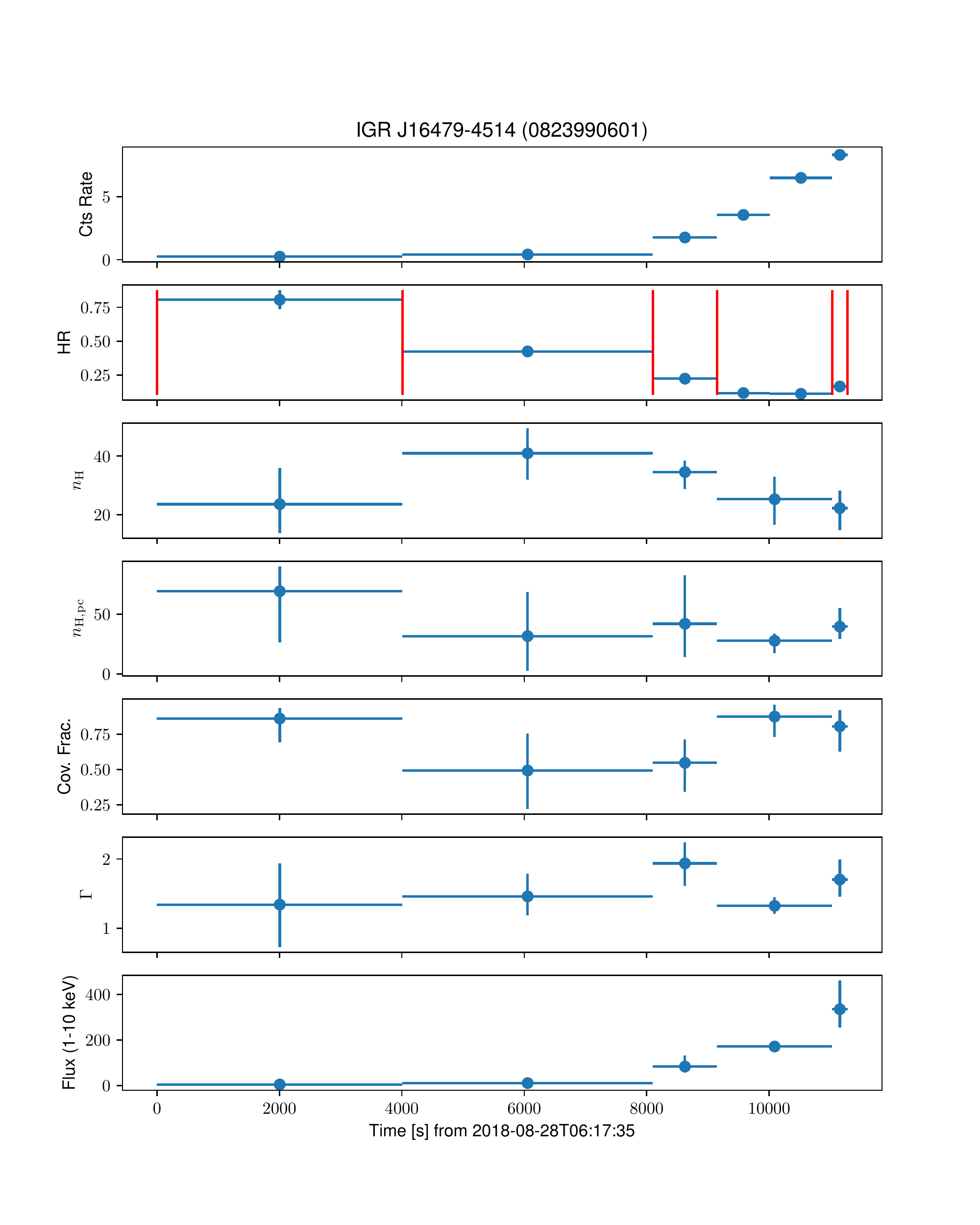}
    \includegraphics[scale=0.45]{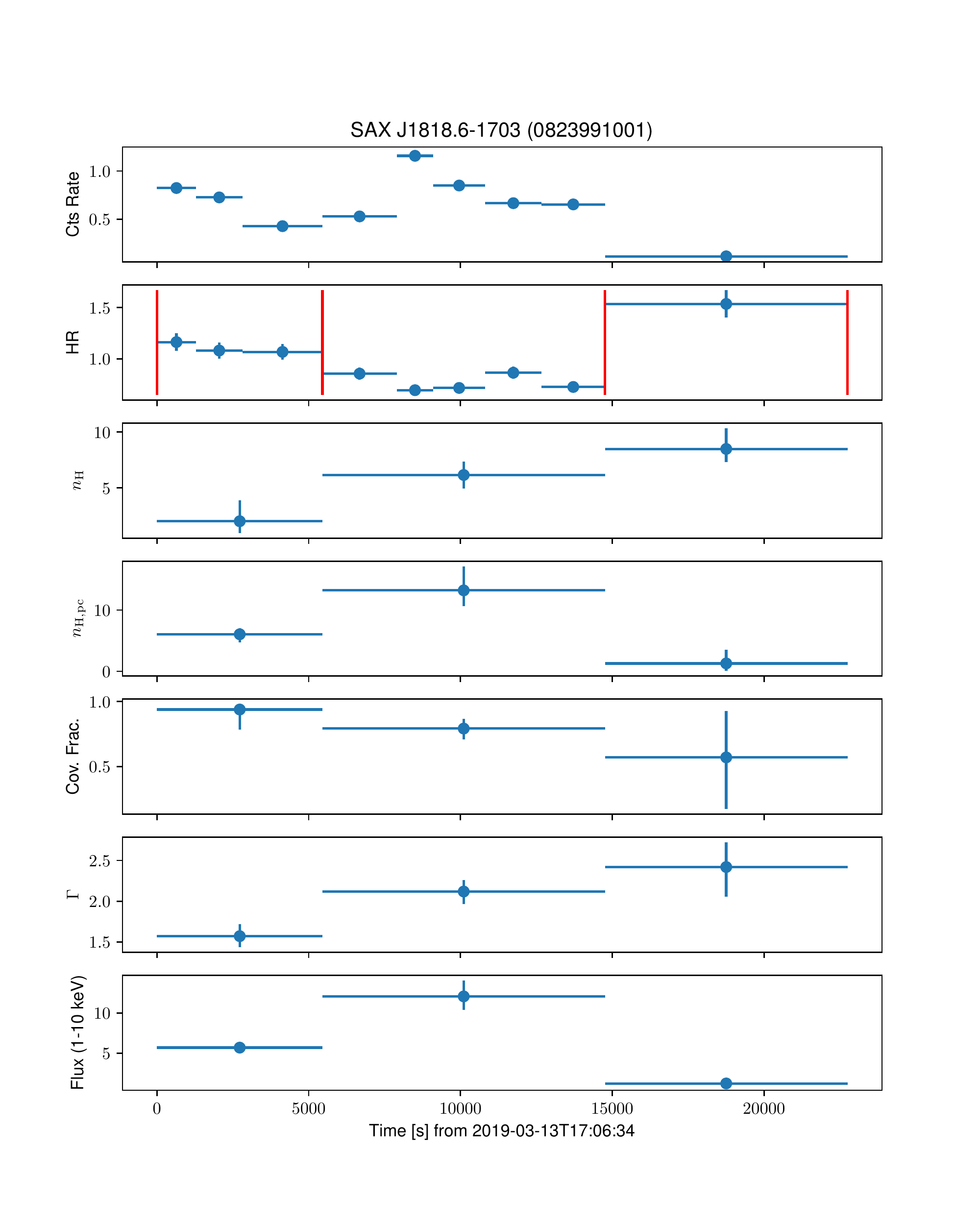}
    \caption{\label{fig:hrspec} In the top panels, for each source, we show the adaptively rebinned lightcurve in the full energy band (0.5--10\,keV). In the second panel from the top, we show the corresponding hardness ratio (HR): the vertical solid red lines indicate the time intervals during which the Bayesian block analysis identifies significant changes in the HR and that were used for the extraction of the HR-resolved spectra. The best-fit parameters with 1-$\sigma$ confidence intervals of the spectra analysis are reported in the panels below (the parameter labeling is the same of Table~\ref{tab:average_spectra}).}
\end{figure*}

\setcounter{figure}{0}
\begin{figure*}
    \centering
    \includegraphics[scale=0.45]{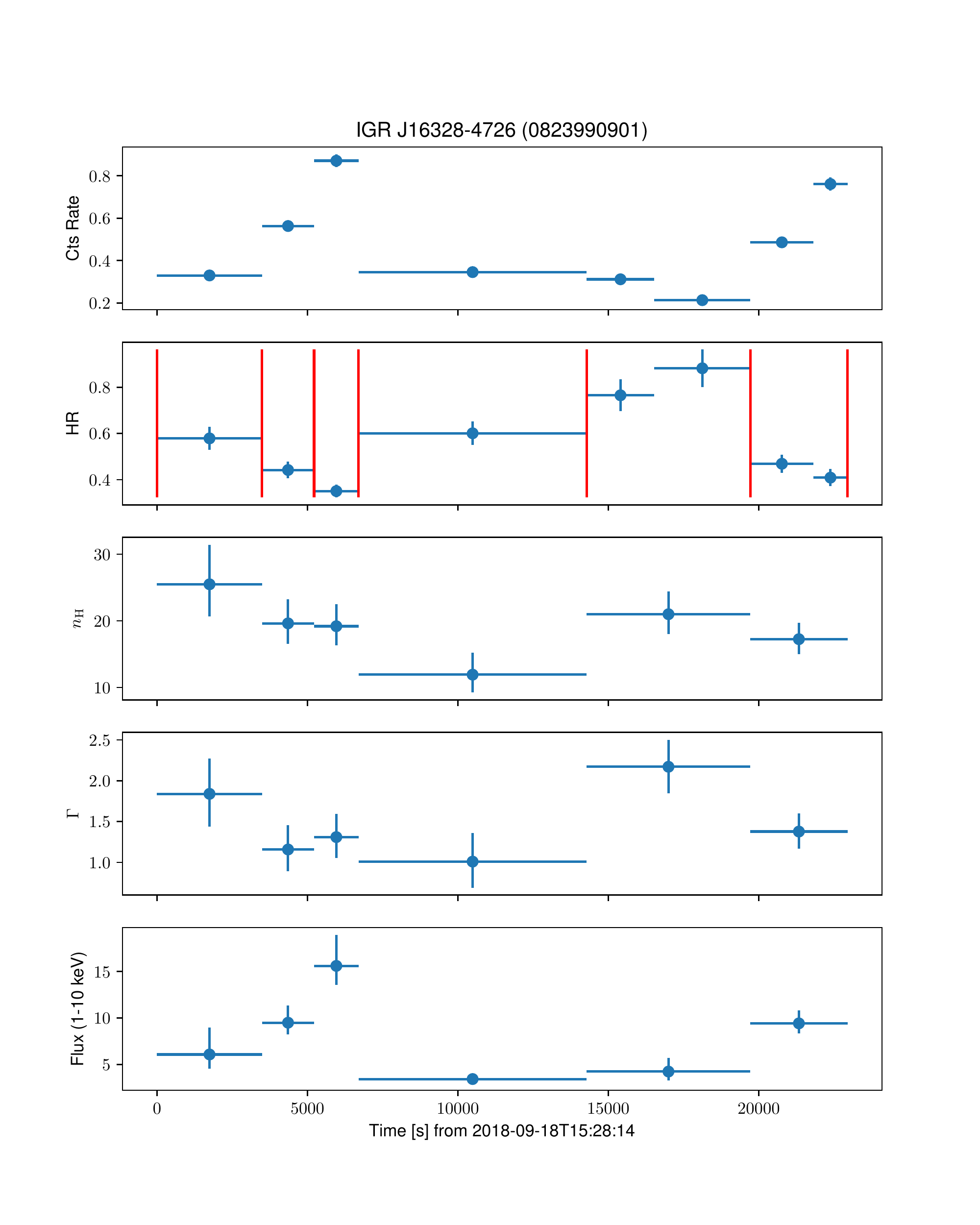}
    \includegraphics[scale=0.45]{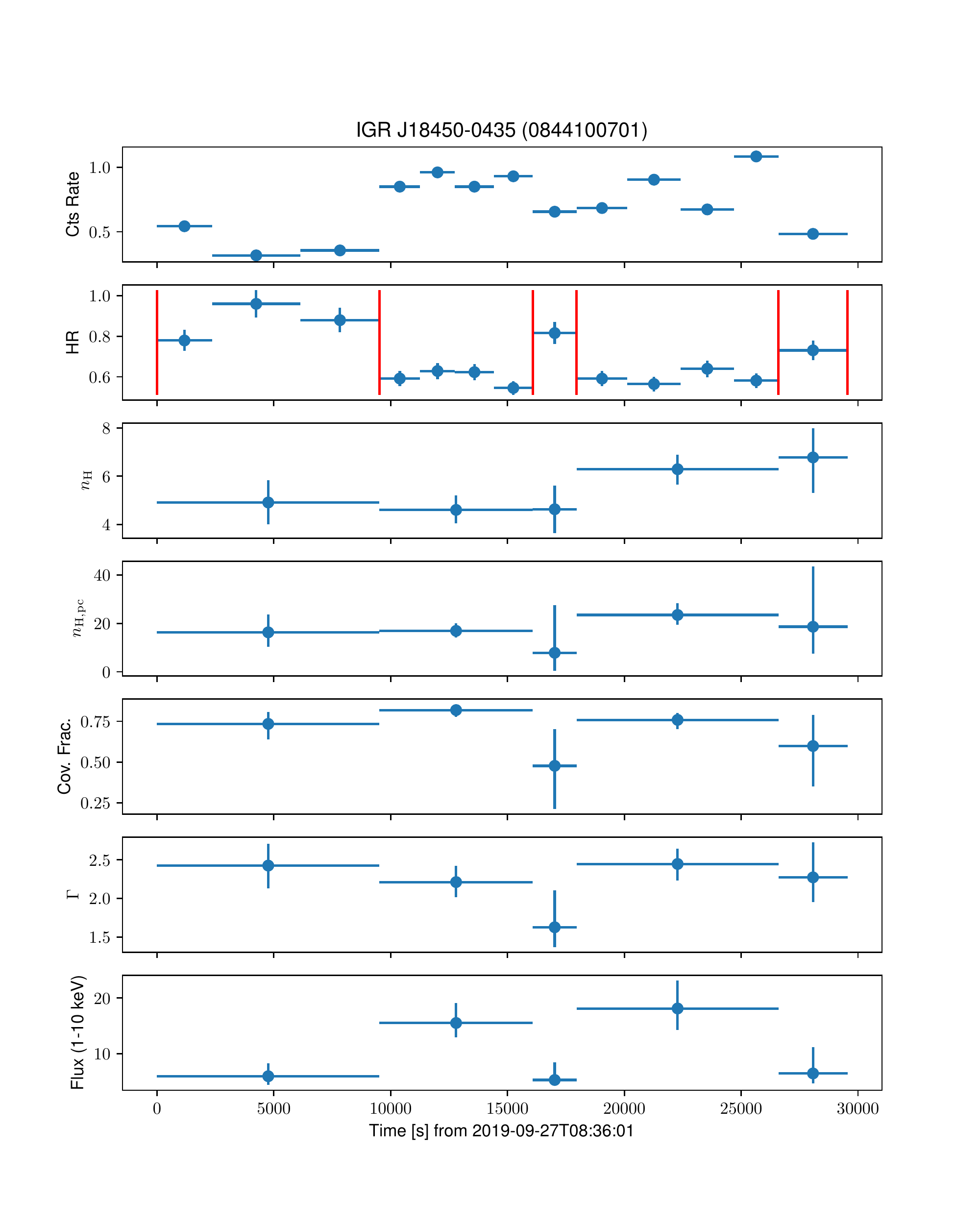}
    \caption{\label{fig:hrspec2} Continued. }
\end{figure*}

To filter out the time intervals corresponding to the flaring background from all data, we extracted the EPIC-pn count rate in the 10--12\,keV energy range in time bins of 100\,s with the standard quality flag \texttt{(\#XMMEA\_EP \&\& PATTERN==0)}. We fitted the count rate histogram of the non flaring part of the lightcurve with a Gaussian. Then, we excluded all 100-s lightcurve bins with probability of less than 0.1\% to belong to the corresponding normal distribution. This filtering criterion resulted in
count rate limits between 0.5 and 1.5 Cts~s$^{-1}$. We visually inspected all resulting lightcurves to check that in this procedure no flares from the sources were erroneously filtered out. As the sources are variable and it is essential for the following analysis to have simultaneous data, we applied the same time selection criteria to all EPIC cameras.

In order to properly select the source and background extraction regions, we first obtained for each observation and each EPIC camera the detector images in the 1--9\,keV energy range using squared image pixels with a size of 80 in detector plane units. As the MOS2 is in all cases operated in timing mode, the image of the central CCD was accumulated in \texttt{RAWX} and time bins of 100\,s. We centered the source extraction region on the best known source coordinates \citep[see][and references therein]{nunez17} and determined its radius as the region within which the average number of photons per image bin is larger than one. This criterion was  chosen based on past experience to optimize the spectral sensitivity (see BZ17). For the EPIC-pn, the background region was in all cases located in the vicinity of the source and characterized by a radius of 1200 detector units (we always visually checked that the background extraction region was free of serendipitous sources, not coincident with traces left on the detector by out of time events, and endowed with a similar stray-light pattern as the target source - if any). For the MOS1 and MOS2, the background extraction regions were placed in the external detectors, as suggested by the standard \xmm\ analysis guidelines\footnote{\url{https://www.cosmos.esa.int/web/xmm-newton/sas-threads}}, and characterized by a radius of 1200 pixels excluding any eventual serendipitous source. In the two observations where the targets were not detected (observation identifier, OBSID 0823990301 and 0844100101, see Table~\ref{tab:observations}), we used a source (background) extraction region with a radius of 550 (2400) pixels to compute a 90\% c.l. upper limit on the source X-ray emission in the 1--10~keV energy band. The size of the extraction regions was chosen large enough to minimize statistical fluctuations in the spectra used to compute the upper limits (see also Sect.~\ref{sec:upper-limits}).

Following the successful approach of BZ17, we searched for variations of the ratio between the source lightcurves extracted in two different energy ranges, i.e. the so called ``hardness ratio'' (HR), in order to highlight the presence of possible spectral variability. The soft lightcurve of each source was extracted between 0.5\,keV and 3--4\,keV, a value that is reasonably close the source photon median energy and the energy at which photo-electric absorption becomes ineffective; the precise upper boundary of the soft lightcurve energy band for each source is reported in Table~\ref{tab:observations} and indicated as $E_\mathrm{HR}$. For each source, we summed the lightcurves of the three EPIC cameras and we adaptively rebinned them adopting a minimal signal to noise ratio (S/N; using the softer one for reference) by accumulating the counts until a pre-defined S/N threshold is reached. The adaptive rebinning was applied starting from the beginning of each observation. The S/N threshold used for most sources is 20, with the exceptions of OBSID 0823990601 and 0823990901, where it was set to 15 due to the lower statistics of the data. In OBSID 0823990801, the source IGR J18450$-$0435 was so weak that even reducing the S/N threshold to 8, we could not evidence any significant variation in the HR. The large variability of the SFXTs across different sources but also for the same source across different observations does not usually make it possible to define a unique S/N for all cases. We verified here, as well as in our previous publications that exploited the same adaptive rebinning method described here, that reasonable variations of this parameter do not affect spectral results, although some fine tuning helps in improving the contrast in spectral variability.

We determined where the HR underwent significant changes by using the Bayesian Blocks analysis \citep{bayesianblocks}, a technique that has been also previously used in the field of SFXTs by \citet{sidoli19} to identify intervals of significant intensity variability in the lightcurves of a sample of these objects observed by \xmm. The block fitness function used is the one for point measurements and the \texttt{ncp\_prior} was chosen to have a false alarm rate around 1\% using Fig.~6 of \citet{bayesianblocks}. We verified that the number of spurious changes of the HR was compatible with expectations by running a sample of 50 simulations with constant count rate drawn from a distribution with noise equivalent to the measured lightcurve of each source. The time intervals identified with this technique where the HR is revealed to undergo significant variations were used to extract simultaneous spectra for the three EPIC cameras for each source (the so-called HR-resolved spectra). The light curve and HR for each source are shown in the two upper panels of Fig.~\ref{fig:hrspec}.
Note that the plot corresponding to the source IGR\,J18462$-$0223 is not reported, as no significant variation of the HR could be revealed.

\begin{table*}
	\renewcommand{\arraystretch}{1.5}
	\caption{\label{tab:average_spectra}Summary of the results obtained from the fits to the average spectra from the \xmm\ observation of each considered source. In the table, $n_{\rm H}$ is the absorption column density in the direction of the source, $n_{\rm H,pc}$ is the absorption column density of the partial covering component, $Cov. Frac.$ is the corresponding covering fraction, and $\Gamma$ is the photon index of the power-law component. We also provide the flux measured in the 1--10~keV energy band (in units of 1$\times$10$^{-10}$~erg~cm$^{-2}$~s$^{-1}$ not corrected for absorption) and the value of the C-statistics together with the degree of freedom of each spectrum.}
	\centering
	\begin{tabular}{cr@{}lr@{}lr@{}lr@{}lr@{}lr@{}lr@{}l}
		\hline
		\hline
		OBSID & \multicolumn{2}{c}{0823990401} & \multicolumn{2}{c}{0823990501} & \multicolumn{2}{c}{0823990601} & \multicolumn{2}{c}{0823990801} & \multicolumn{2}{c}{0823990901} & \multicolumn{2}{c}{0823991001} & \multicolumn{2}{c}{0844100701}\\
		\hline
		$n_\mathrm{H}$ & 9.0 &$\pm$ 0.5 & 3.3 &$\pm$ 0.6 & 18 &$\pm$ 5 & 27 &$\pm$ 6 & 19 &$\pm$ 1 & 3.5 &$\pm$ 0.8 & 4.0 &$\pm$ 0.4  \\
		$n_\mathrm{H, pc}$ & 21 &$\pm$ 1 & 7.5 &$\pm$ 1.9 & 36 &$\pm$ 3 & --&-- & --&-- & 9.2 &$\pm$ 0.9 & 13 &$\pm$ 2  \\
		Cov. Frac. & 0.71 &$\pm$ 0.03 & 0.77 &$\pm$ 0.09 & 0.90 &$\pm$ 0.06 & --&-- & --&-- & 0.90&$_{-0.05}^{+0.04}$ & 0.74 &$\pm$ 0.03  \\
		$\Gamma$ & 1.27 &$\pm$ 0.04 & 2.4 &$\pm$ 0.2 & 1.4 &$\pm$ 0.1 & 2.4&$_{-0.2}^{+0.3}$ & 1.5 &$\pm$ 0.1 & 1.88 &$\pm$ 0.10 & 1.96 &$\pm$ 0.10  \\
		Flux (1-10 keV) & 90 &$\pm$ 3 & 2.0&$_{-0.3}^{+0.4}$ & 46 &$\pm$ 5 & 1.1&$_{-0.1}^{+0.4}$ & 6.7 &$\pm$ 0.5 & 5.9 &$\pm$ 0.4 & 8.2 &$\pm$ 0.7  \\
		Cstat/d.o.f. & 305&/272 & 264&/242 & 241&/218 & 326&/234 & 165&/135 & 254&/244 & 302&/253  \\
		\hline
	\end{tabular}
\end{table*}

In order to determine the best spectral model to be used for the HR resolved spectra, we extracted the average spectrum of each source by integrating over the entire exposure time available in each observation.  Being endowed with the best achievable statistics for a specific observation, these total spectra allow us to find the most appropriate description of the source X-ray emission in the \xmm\ energy band (this is the same technique that we adopted in a number of our previous papers on the SFXTs \citep[see, e.g.,][BZ17]{bozzo11,bozzo15}. The spectral analysis was performed in all cases with \texttt{Xspec} version 12.10.1f \citep{xspec} using the W statistic (\texttt{cstat})) after optimally grouping each source spectrum as described in \citet{Kaastra2016}. We excluded from all fits data below 0.55 keV and above 10 keV for the EPIC pn and MOS1, while MOS2 data were discarded also between 0.55 keV and 2 keV because of clear instrumental residuals in this energy band in contrast with those observed in the simultaneous MOS1 and EPIC pn spectra (but see also deviations in Sect.~\ref{sec:deviations}).

Based on our long-standing expertise on the analysis of SFXTs with \xmm,\ we first attempted to fit the time-averaged spectra of all sources with a simple absorbed power-law model. This model provided acceptable results only in the cases of IGR\,J18462$-$0223 and IGR\,J16328$-$4726. In all other cases, the simple absorbed power-law model left evident structures in the residuals from the fits at both the lower and upper energy bounds covered by the EPIC cameras. We could significantly improve the fits for almost all these sources (the only exception being  IGR\,J18483$-$0311, see later in this section) by adopting a power-law model  obscured by two layers of neutral absorption, one covering completely the source (\texttt{TBabs} in \texttt{Xspec}) and the other partially extinguishing it (\texttt{pcfabs}). Abundances were set to \texttt{Wilm} \citep{Wilms2000} and cross section to the values reported in \citet{Verner1996} for all cases, following the common approach for SFXT sources and HMXBs in general \citep[see, e.g.,][and references therein]{walter2015}. Cross calibration uncertainties and data selection effects induced by the instrument good time intervals were accounted for in all cases by introducing cross-calibration constants in the fits. These all turned out to be close to unity. 

\begin{table}
	\renewcommand{\arraystretch}{1.5}
	\caption{\label{tab:average_spectra_02}Same as Table~\ref{tab:average_spectra} but here the results are shown in the specific case of IGR\,J18483$-$0311 which is the only source requiring the addition of an APEC spectral component to acceptably fit its X-ray spectrum.}
	\centering
	\begin{tabular}{cr@{}l}
		\hline
		\hline
		OBSID & \multicolumn{2}{c}{0823990201} \\
		\hline
		$n_\mathrm{H, APEC}$ & 2.0 &$\pm$ 0.2  \\
		$kT_\mathrm{APEC}$ & 0.16 &$\pm$ 0.02  \\
		$N_\mathrm{APEC}$ & 0.031&$_{-0.014}^{+0.021}$  \\
		$n_\mathrm{H}$ & 6.4 &$\pm$ 0.3  \\
		$n_\mathrm{H,pc}$ & 13 &$\pm$ 1  \\
		Cov. Frac. & 0.60 &$\pm$ 0.03  \\
		$\Gamma$ & 1.63 &$\pm$ 0.03  \\
		Flux (1-10 keV) & 98 &$\pm$ 2  \\
		Cstat/d.o.f & 441&/317\\
		\hline
	\end{tabular}
\end{table}

For all fits, we first performed a preliminary minimization using the Levenberg-Marquardt algorithm based on the \texttt{CURFIT} routine from Bevington. Then, we explored the space of each free parameter using a Monte Carlo Markov Chain (MCMC) generated by the Goodman-Weare algorithm with 40 walkers, a length of 26\,000, and a burning phase length of 6000. We adopted Jeffreys priors for the unabsorbed power-law flux and absorption column densities. The latter were constrained within the limits of $10^{21}$ and $10^{24}\, \mathrm{cm}^{-2}$. We used  linear priors for the covering fraction and power-law photon index, limited in the intervals 0--1 and -1--6, respectively. From the posterior distributions, we determined  the equivalent 1-$\sigma$ uncertainties on each fit parameter for each source using the 16 and 84\% percentiles. To test the statistical goodness of our model, we took 100 parameter realizations from the MCMC and simulated the EPIC spectra with the exposure time and the background count-rate equivalent to the measured ones. We then computed the fit statistics and compared its distribution with the best-fit value. Generally, we found that from 20 to 60\% of the simulated spectra were characterized by a C-stat larger than the best-fit C-stat. There are two exceptions to this finding. The first concerns OBSID 0823990901, for which the time-averaged spectrum is noisy at low energy owing to background fluctuations. The second concerns OBSID 0823990201, which time-averaged spectrum showed convincing evidences for an additional spectral component below $\sim$2~keV (see below). We also note that the Bayesian posterior sampling could be sensitive to the particular algorithm used to explore the parameter space, and thus we also verified that the posteriors determined with the Goodman-Weare algorithm were consistent with the ones determined using the multinest method \citep{multinest} in each average spectrum. For this task, we exploited the BXA interface to \texttt{Xspec} \citep{bxa}. Results were in all cases consistent (within uncertainties), so we used and reported in the following only the results obtained with the Goodman-Weare algorithm.

The results of this analysis are shown in Table~\ref{tab:average_spectra}.
For all these sources, the same model was also used to fit the HR-resolved spectra and all results (best fit spectral parameters vs. time) are shown in Fig.~\ref{fig:hrspec}. Note that in the case of the source IGR\,J18462$-$0223 only the time averaged spectrum was extracted due to the low statistics of the data and the lack of any significant HR variability.

The only source which averaged and HR-resolved spectra could not yet be satisfactorily fit with the partial absorbed model is IGR\,J18483$-$0311 (OBSID 0823990201). In the case of this source, the fit with the partial absorbed model left evident residuals at energies $\lesssim$1.8~keV.
We attempted to improve the fit changing the partial absorbing component with any other physically motivated component usually adopted for SFXTs (e.g., black body, disk black body, breemstrahlung, additional power-law). None of these attempts resulted in an acceptable fit. We thus concluded that a further spectral component was needed beside those already included in our partially absorbed model. We found that the addition of a component likely emerging from the wind of the supergiant star \citep[see, e.g.,][]{bozzo10, sidoli10} and originated from a plasma in ionization equilibrium (APEC model in {\sc Xspec}) could remove the most significant structures in the fit residuals (see Fig~\ref{fig:spec_02}). We summarize all results obtained from the fit with this model to the IGR\,J18483$-$0311 data in Table~\ref{tab:average_spectra_02}.

We verified that the parameters of the APEC component could not be constrained in the fits to the HR-resolved spectra of IGR\,J18483$-$0311, and thus we kept in these cases both the APEC temperature and column density fixed to the values measured from the time averaged spectrum. Only the normalization of the APEC component was left free to vary in the HR-resolved spectra (see Fig.~\ref{fig:spec_res_0823990401} for the best-fit parameters).

\setcounter{figure}{1}
\begin{figure}
	\centering
	\resizebox{\hsize}{!}{\includegraphics{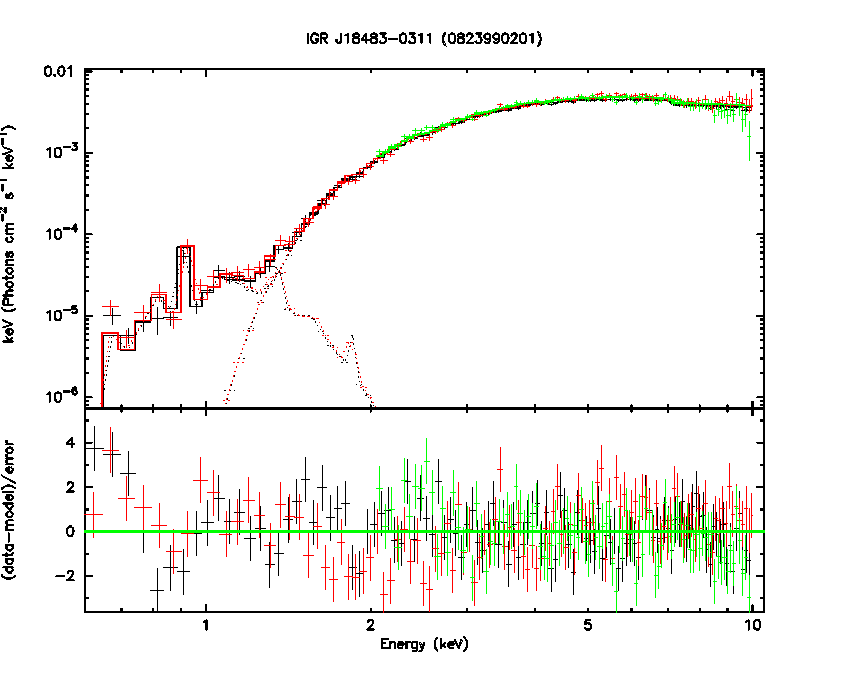}}
	\caption{\label{fig:spec_02}Unfolded spectrum of IGR\,J18483$-$0311, together with the best-fit model summarized in Table~\ref{tab:average_spectra_02}. The residuals from the fit are shown in the lower panel. Data points and lines in black correspond to the EPIC-pn data, while those in red and green represent the MOS1 and MOS2 data, respectively.}
\end{figure}

\begin{figure}
    \centering
    \resizebox{\hsize}{!}{\includegraphics{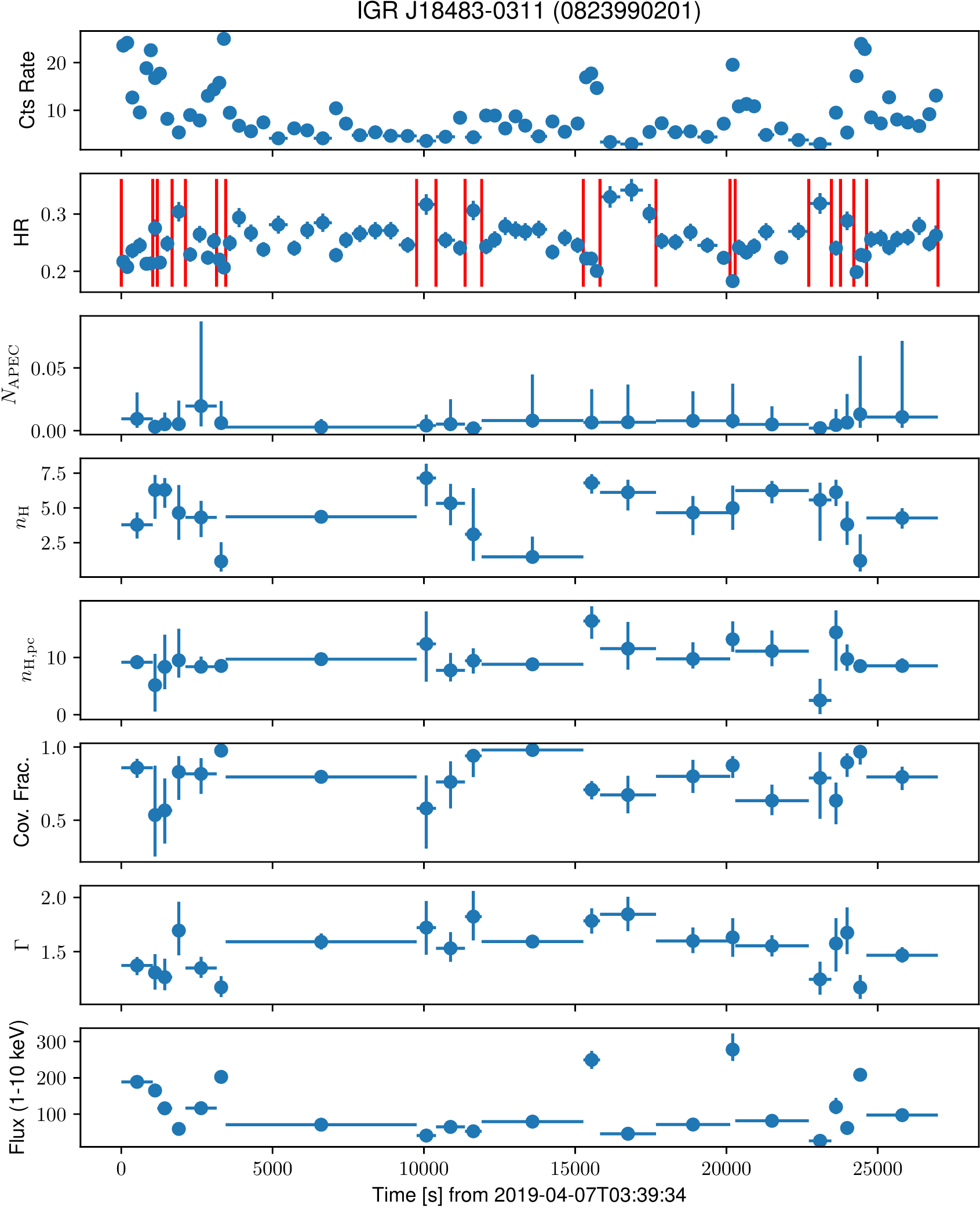}}
    \caption{\label{fig:spec_res_0823990401} Same as Fig.~\ref{fig:hrspec} but in the case of the source IGR\,J18483$-$0311, which is the only one requiring the addition of an APEC component to satisfactorily fit its time-averaged spectrum (see text for more details).}
\end{figure} 

We have performed a systematic search for correlations between the best fit parameters obtained from the HR-resolved spectra of all analyzed sources. In each source, the linear correlation was computed between two fit parameters taking into account uncertainties in both variables and extracting their values from a symmetric Gaussian distribution with standard deviation determined from half the 16--84\% posterior interval. We could not find any significant correlation, the only exception being the obvious case of the source IGR\,J16418$-$4532, where the covering fraction appears to be  clearly anti-correlated  with the count rate (see Fig.~\ref{fig:correlation}). In this case, the Person's coefficient 1-$\sigma$ interval in our bootstraped realization is contained between 0.48 and 0.62, while the slope is $-0.14\pm0.02$ at 68\% c.l.

\subsection{Optimization of our standard analysis}
\label{sec:deviations}

We list in this section a few deviations compared to the description of the general data reduction and analysis provided in the previous section.
Although the main analysis steps are the same for all observations, as illustrated below, some data-sets required some ad-hoc adjustment
of the general procedure to avoid any spurious alteration of the spectral fit outcomes. This has been a common practice also
for our previous publications on the subject (see, e.g., BZ17).
For OBSID 0823990201, we excluded data after 2019-04-07 at 11:10:56 because a particularly bright solar flare made the data unusable from this time until the
end of the planned exposure. No other filtering was required in the remaining part of the observation.
For this observations, the somewhat lower absorption than in other cases (see Sect.~\ref{sec:results}) allowed us to use also MOS2 data
in the energy range 0.55-2~keV as no anomalous residuals were found compared to the other EPIC cameras.

For OBSID 0823990401, we excluded EPIC-pn data up to 2\,keV and MOS1 from 1.2 to 1.5\,keV, due to evident background-induced residuals.

For OBSID 0823990601, we used for the scientific analysis only data collected from 2018-08-28 at 10:47:44 onward, i.e. 16\,ks after the nominal start of
observation. This was necessary in order to exclude an initial time interval in which the source was too weak to obtain meaningful results. Only a few minor
solar flares were observed in the following period, resulting in a nearly negligible good time interval filtering.
The lower energy bound of EPIC-pn data for the spectral analysis is set to 2\,keV, due to residuals in the fits not compatible with those of the MOS1,
indicating a clear instrumental origin. For the same reason, MOS2 data were excluded below 3\,keV.

As known from previous publications \citep{zurita09}, IGR\,J18450$-$0435 is located close to a known Active Galactic Nucleus. Thus, in OBSID 0844100701, we limited the source extraction region to 640 pixels in imaging mode and to a width of 32 \texttt{RAWX} units for timing mode.

In the case of OBSID 0823990901, MOS1 was operated with a closed filter for virtually all available observational time
and no useful data could be exploited for the scientific analysis. Due to instrumental residuals, the MOS2 data below 3~keV were ignored.

\subsection{Computation of flux upper limits}
\label{sec:upper-limits}

During OBSID 0823990301 and 0844100101, the targets were not detected by \xmm.\
To compute the upper limits on their X-ray emission, 
we searched in the literature for the most common spectral model used to describe the X-ray emission from IGR\,J11215$-$5952 and IGR\,J18410$-$0535 in an energy band compatible with that of the \xmm\ EPIC cameras. 
For IGR\,J18410$-$0535, we adopted the absorbed-power-law model used by \citet{bozzo11} in the first bin of their lightcurve. At the best of our knowledge, this is the most accurate description of the source spectrum at low luminosity.
For IGR\,J11215$-$5952, we adopted the partial covering model for the average spectrum reported in \citet[][their model 2 in Table 1]{Sidoli2017}. We set Gaussian priors to all model parameters except the integrated flux of the power-law in the 1--10\,keV band, for which we use a Jeffrey's prior. The Gaussian average is the literature best-fit parameters, while the standard deviation is the average of the upper and lower uncertainties scaled at 68\% c.l. We used data for EPICpn only, as it is critical to estimate the background very close to the source and this is better achieved in full window mode. After running the Goodman-Weaver MCMC chain in Xspec, we determined the upper 90\% percentile on the normalization as the unabsorbed flux upper limit. 
The results are reported in Table~\ref{tab:observations}. For the MOS cameras, uncertainties linked to background subtractions far from the source extraction region prevent a meaningful upper limit estimation.

\section{Results}
\label{sec:results}

The lightcurves of the different sources displayed in the top panels of Fig.~\ref{fig:hrspec} 
show that our observational program has been successful as up to several flares are detected in most of the $\sim$20~ks long \xmm\ pointings. This was expected based on our current knowledge on the SFXTs from literature results and long-term studies carried out especially with \swift/XRT. The only exceptions are the sources IGR\,J11215$-$5952 and IGR\,J18410$-$0535, which were not detected during the \xmm\ observation, as well as the source IGR\,J18462$-$0223, which was caught in a low activity state. We comment below individually on the results obtained for each source included in the present study.

IGR\,J18483$-$0311: our \xmm\ observation found the source in a remarkably active state and a total of at least six flares were recorded in less than 30\,ks. Some of these flares achieve a count-rate as high as $\sim$25~cts~s$^{-1}$ in the EPIC-pn camera (see Fig.~\ref{fig:spec_res_0823990401}).
Although the observed flares are bright, the variations of the HR are relatively limited (less than a factor of two). Looking at the relevant plot in Fig.~\ref{fig:spec_res_0823990401},
we note that the lowest HR values are recorded in between flares around 10\,ks, 12\,ks, 17\,ks, and 24\,ks from the beginning of the observation. The analysis of the HR-resolved spectra revealed {\bf hints of a} lower covering fraction during the low HR time intervals (especially for the interval around 24~ks) and a correspondingly increased absorption column density. There seems to be some evidence for a larger covering fraction during the rises and peaks of the flares, although the statistics is not sufficiently high to reach a firm conclusion (throughout the paper we refer to rises and decays of the SFXT flares following the same definition and approach used in our previous paper BZ17). The normalization of the APEC component was in general poorly constrained and we could not find indications of any significant variability during the selected time intervals for the spectral analysis. It should, however, be noted that including this component in all fits avoided any bias during the comparison of the results obtained from different time intervals.

\begin{figure}
	\centering
	\resizebox{\hsize}{!}{\includegraphics{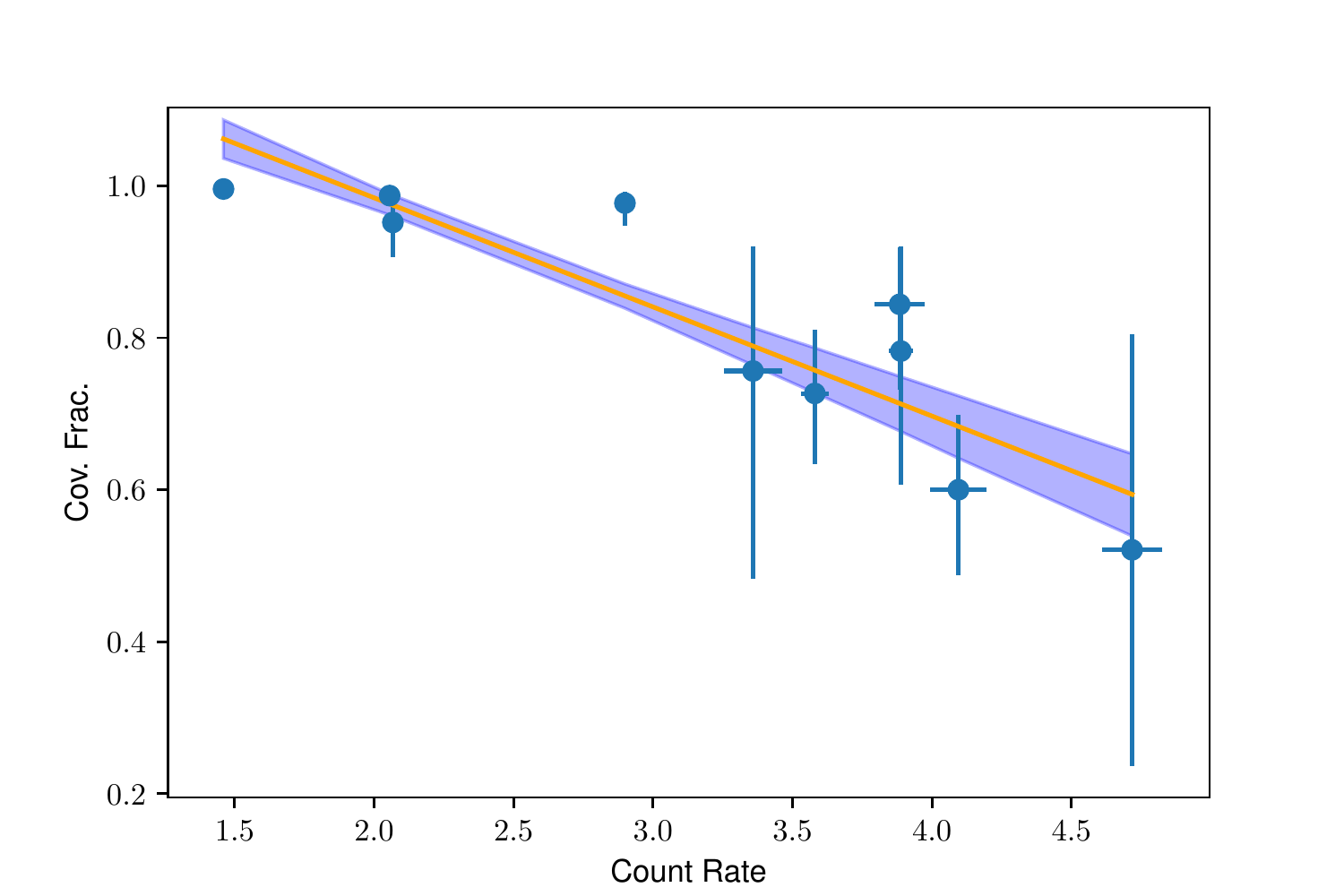}}
	\caption{\label{fig:correlation}Correlation between the covering fraction and the source count-rate in IGR\,J16418$-$4532 (OBSID 0823990401). The shaded region indicated the envelope of the correlation curves at 1$\sigma$ confidence level.}
\end{figure}

IGR\,J16418$-$4532: among the observations presented in this paper, the OBSID 0823990401 is certainly the most interesting. Looking at the relevant plot in Fig.~\ref{fig:hrspec}, we can see that the source underwent at least five distinct flares and displayed a progressively decreasing HR which varied by a factor of$\sim$7 between the beginning and end of the observation. Apart from the overall progressive decrease, the recorded HR shows significant increases at the flare rises and abrupt decreases toward the flare peaks.
The results of our spectral analysis reported in Fig.~\ref{fig:hrspec} confirm that the progressive change in the HR is due to a variation of the column density. This analysis  shows that the increases of the HR during the rises to the flares are also associated to increases in the absorption column density, while the drops of the HR during the peaks of the flares are associated with both a decrease in the absorption column density and in the covering fraction. This conclusion is further confirmed by the detection of a significant anti-correlation between the source count-rate and the covering fraction displayed in Fig.~\ref{fig:correlation} and introduced earlier in Sect.~\ref{sec:data_analysis}.

\begin{figure}
	\centering
	\resizebox{\hsize}{!}{\includegraphics{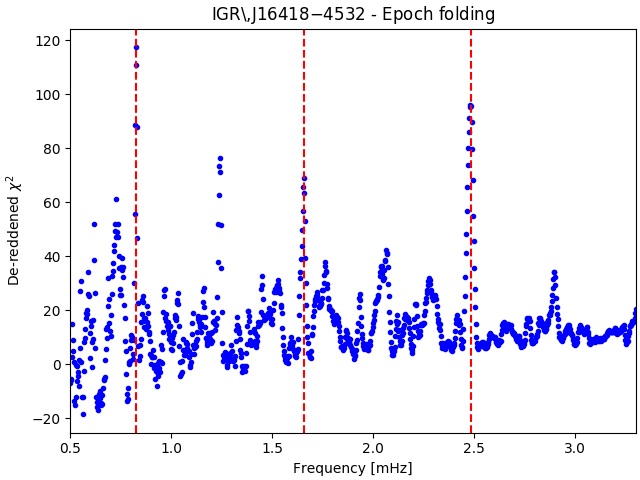}}
	\caption{\label{fig:epoch_folding}The periodogram of IGR\,J16418$-$4532 (OBSID 0823990401) obtained with epoch folding using 16 phase bins. Red noise has been modelled and subtracted. The red vertical dashed lines correspond to the pulse frequency of 0.827 mHz and its two harmonics.}
\end{figure}

We note that these variations cannot be related to the pulsations from the source as the latter occur recurrently on a much shorter time scale
\citep[$\sim$1210~s; see, e.g.,][and references therein]{drave13}.
To search for pulsation in our data set,
we performed an epoch folding search \citep{Leahy1987} on the combined 0.5--10\,keV light curve of the source with 10\,s bins.
We explored frequencies from 0.5 to 3.3 mHz with no weighting using 16 phase bins.
The lower limit is chosen to exclude the secular trend from the periodogram red-noise and the upper is determined from the light curve binning.
To subtract the red noise, we fit a power law to the frequency dependency of the $\chi^2$ and found a slope of $0.99\pm0.03$
with normalization at 1 Hz of $3.1\pm0.5\times10^{-3}$.
We subtract this function and add the expectation value of a $\chi^2$ distribution
with 15 degrees of freedom and obtain the curve of Fig.~\ref{fig:epoch_folding}.
The most prominent peak is at a frequency of 0.827$\pm$0.003\,mHz, corresponding to a period of 1208$\pm$4\,s; two other peaks are
present at twice and three times the base frequency. We consider the peak at 1.25\,mHz as spurious.
Following \citet{Dai2011}, we determine the period uncertainty as $P_\mathrm{max}/2 N_\mathrm{ph} \Delta t$, where
$P_\mathrm{max}$ is the peak frequency,  $\Delta T$ the observation duration, and $N_\mathrm{ph}$ the number of trial phase bins.
To asses the significance of the peak, we compute an exponential fit to the histogram of de-redenned $\chi^2$ between
the values 20 and 50 using an exponential function ($K\exp(-b\chi^2)$). We find $K=230\pm30$ and $b=0.104^{+0.006}_{-0.003}$ at 68\% c.l.
Then, we compute the 90\% quantile of the integrals of the normalized exponential fitting function above the $\chi^2$ value of the peak.
Converting this probability into standard deviations of a normal distribution,
we find that the main peak has an equivalent significance of 3.7$\sigma$.
If we compute the combined probability that also the harmonics are present, we reach
a robust detection of the pulsation at $5\sigma$.  
Our detection of the spin period in this observations is in line with previous findings
in \citet{sidoli12, drave13} and does not yield any measurable spin evolution.

AX\,J1949.8+2534: this source was caught by \xmm\ in a relatively faint state, but we recorded an overall variation in the HR by a factor of $\sim$2.5 across the observation. Although the statistics is limited, the results of the HR-resolved spectral analysis in Fig.~\ref{fig:hrspec} suggests a similar situation as reported for the previous source, with an overall decrease of the absorption column density during the observation. There is a hint for a faint short flare about 13~ks after the beginning of the observation but the event lasted for a too short amount of time (about 1~ks) and was too faint to perform any meaningful more detailed investigation.

IGR\,J16479$-$4514: this source displayed a remarkable increasing HR, reaching a value about 10 times higher from the beginning up to the first 12~ks of the observation and undergoing an abrupt decrease by a factor of $\sim$2 at the very end of the observation (the last ks; see Fig.~\ref{fig:hrspec}). The progressive increase in the HR together with the correspondingly increase in the source count-rate strongly remembers what is usually observed during the egress from an X-ray eclipse. IGR\,J16479$-$4514 is known to display X-ray eclipses \citep[see, e.g.,][]{bozzo09}, and if we consider the most recently published ephemeris of the source \citep{coley15}, it is possible to show that the \xmm\ observation studied in this paper is compatible with being the egress from an eclipse. The increase in the source count-rate across the \xmm\ observation is thus likely associated with the source ramping up to its usual emission state and it is not a flare. It remains puzzling the sudden drop in the HR at the very end of the observation. Results of the HR-resolved spectral analysis do not show a clear trend and it is likely that the progressive increase in the HR is the combined effect of slight variations in several parameters (absorption column density, covering fraction, power-law photon index). There is some evidence favoring a decrease of the absorption column density and increase in the covering fraction to explain the abrupt decrease of the HR in the last ks of this observation.

SAX\,J1818.6$-$1703: this source underwent two relatively faint flares during the \xmm\ observation (Fig.~\ref{fig:hrspec}). Although the statistics is far too low to perform a detailed study of the HR variations within the rises/decays of the two flares, our HR-resolved spectral analysis (Fig.~\ref{fig:hrspec2}) revealed a progressive increase in the absorption column density during the entire observation, accompanied by a softening of the power-law and a marginally significant decrease of the covering fraction. Although the source ephemeris have been reported in the literature \citep{bird09}, we have been unable to determine the orbital phase of the present \xmm\ observation. This is due to the relatively large uncertainty associated with the orbital period in the available parameters and the extrapolation of the orbital phase up to 2019.

IGR\,J16328$-$4726: this source underwent two flares during the observation and displayed a relatively limited but interesting variation of the HR (see Fig.~\ref{fig:hrspec2}). The HR is observed to increase toward the peak of the flares and decrease between them. Our HR-resolved spectral analysis revealed that the increase is the result of two opposite effects. On the one hand, the absorption column density rises at the onset of the flare and decreases toward the peak. On the other hand, there is a substantial hardening of the power-law photon index from the onset to the peak of the flare that compensate the variation of the absorption column density and contributes to make the overall source spectral emission harder at the peaks of both recorded flares. The emission during the quiescent time interval between the two  flares is characterized by a low HR which results uniquely from the low absorption column density (the lowest measured during the entire observation) as the power-law photon index is the hardest measured across the $\sim$25~ks spanned in the OBSID~0823990901.

IGR\,J18450$-$0435: this source underwent three faint flares during the $\sim$20~ks long \xmm\ observation. Although the statistics was virtually identical to that of the source SAX\,J1818.6$-$1703, the overall variation of the HR was slightly less pronounced  (Fig.~\ref{fig:hrspec}) and our HR-resolved spectral analysis could not identify any significant spectral change (albeit all measured values of the spectral parameters are endowed with fairly large error bars, see Fig.~\ref{fig:hrspec2}).

\section{Discussion and conclusions}
\label{sec:discussion}

This paper reports the results obtained from the first ten observations of our on-going monitoring program of the SFXTs with \xmm\ and serve as a demonstration that our program, albeit started only recently, is successfully delivering the expected outcomes. The program is aimed at carrying out roughly 20~ks-long observations in the direction of all known SFXTs to populate the database of flares observed from these sources with the only X-ray facility that is endowed at present with the right combination of sensitivity and spectral resolution to detect fast spectral variations during these events. As discussed in the literature, such studies might ultimately help us in understanding the mechanism(s) driving the peculiar behavior of the SFXTs in X-rays (see also Sect.~\ref{sec:intro}). So far, most of the already performed 20~ks-long pointings caught from one to six flares from the targeted SFXTs.

The richest dataset acquired so far, and certainly the most intriguing is the one for the SFXT IGR\,J16418$-$4532. This source displayed a progressively decrease in the absorption column density along the entire \xmm\ observation plus significant swings of the same spectral parameter during the rise and decay from the flares\footnote{Note that our \xmm\ observation did not occur during the eclipse of the source as the time of the observation is incompatible with the expected times according to the ephemeris reported by \citet{coley15} also when all uncertainties are taken into account.}. This behavior is remarkably similar to that observed in the case of the SFXT IGR\,J17354$-$3255 during the \xmm\ OBSID 0693900201 (see Fig.~9 in BZ17).  Following our previous interpretation of the event from IGR\,J17354$-$3255, we also suggest that in the case of IGR\,J16418$-$4532 \xmm\ might have observed a relatively rare case in which a large massive clump has passed in front of the neutron star along the line of sight to the observer. Such event causes a substantial obfuscation of the X-ray source due to the increased local absorption column density and the flares that are observed during the pointing are likely triggered by some part of this possibly structured clump onto the NS. Interestingly, the reported increases in the absorption column density during the rises of flares and the abrupt decreases close to the peaks are similar to what has been observed in the past in several SFXTs (BZ17) and interpreted as being due to the approaching of a clump (or some structures within it) to the neutron star (during the rises from the flare) and the photoionization of the clump material shortly after the flare has reached a sufficiently high luminosity (close to the peak). The significant anti-correlation between the count rate and the absorption covering fraction is a further indication that an increasing radiation tends to clear up wind clumps around the neutron star. These trends in the spectral variability support the idea that clumps play a major role in triggering the flares/outbursts from SFXTs but, as commented in previous literature papers, it cannot be excluded that additional mechanisms are at work to inhibit accretion in SFXTs for most of the time and permit accretion only when the increase in the local mass accretion rate caused by the clump is hampering their effectiveness in controlling the accretion flow \citep[see, e.g., the discussion in][and references therein]{bozzo15}.

In the past, IGR\,J16418$-$4532 has also shown other episodes of intriguing X-ray activity. During another \xmm\ observation, \citet{drave13} reported about a large temporary increase in the local absorption column density that lasted about 1~ks and occurred slightly before the rise of a flare. This occurrence was also interpreted in terms of a clump approaching the neutron star and then been accreted onto the compact object (giving rise to the subsequent X-ray flares). In 2011, a peculiar episode of unusual low variability emission was observed with \xmm\ and ascribed to the possible switch from the usual wind accretion regime to a Roche-Lobe overflow regime, during which an accretion disk is suspected to form around the neutron star\citep{sidoli12}. This conclusion remained speculative, given the lack of direct evidences for the presence of an accretion disk, as well as the lack of other similar events during the later observations of the source. Although the dynamic range of the X-ray luminosity displayed by IGR\,J16418$-$4532 is somewhat on the low side compared to most of the SFXTs \citep[see also the discussion in][]{bozzo15}, the observational findings on this source make it one of the most promising candidates to search for spectral variability during and in between flares.

A number of other sources reported in this paper showed some similarity in their spectral variability during flares with IGR\,J16418$-$4532, especially for what concerns the changes in the absorption column density. Of particular interest for the goal of our analysis is the detection of enhancements in the absorption column density just before or during the rise of a flare, as well as the decrease in the $n_{\rm H}$ at the peak of the flares. As summarized in Sect.~\ref{sec:intro} and briefly mentioned earlier in this section, similar indications support the idea of clumps being key players in driving the variability of SFXTs. In Sect.~\ref{sec:results}, we showed that evidences for a similar behavior could be obtained for IGR\,J16328$-$4726 and, albeit with more uncertainties due to the low statistics, in the source AX\,J1949.8+2534. IGR\,J16479$-$4514 showed an evidence of a decreasing absorption column density close to the peak of a possible flare, although this result has to be taken with caution because \xmm\ most likely caught the source while emerging from an X-ray eclipse.

For the remaining sources, the interpretation is less clear. IGR\,J18483$-$0311 displayed several bright flares but despite the highest count-rate achieved compared to all other sources presented here, the overall variation of HR remained relatively low. We could neither reveal the expected rises of the $n_{\rm H}$ close to the onset of the flares, nor the drops around their peaks. We found evidence of a larger covering fraction during the most intense X-ray luminosity time intervals and lower values of the same parameter during the times between the flares. In the context of clumpy wind interpretation, we could argue that during this specific \xmm\ pointing the source was located in a particularly dense region of the wind and flares were triggered by just modest variations of the local mass accretion rate likely overwhelming the effect of the mechanisms usually inhibiting accretion. The higher covering fraction during the flare might indeed be connected to the slightly larger amount of accreting material from the stellar wind getting closer to the neutron star. A similar scenario could be applicable in the case of SAX\,J1818.6$-$1703, which showed a tentative evidence of a decrease in the covering fraction during the lowest emission time interval toward the end of the \xmm\ observation. The soft APEC component revealed in the spectrum of IGR\,J18483$-$0311 could not be studied in much detail, due to the lost statistics at energies $\sim$2~keV (especially its variation as a function of time and/or HR). However, its detection is already quite interesting because it likely indicates the presence of a strong stellar wind, as suggested in the case of a similar soft spectral component revealed in the X-ray emission of the SFXT IGR\,J08408$-$4503 \citep{bozzo10,sidoli10}. Should shocks be present in supergiant star winds,
as predicted by one-dimension numerical simulations \citep{feldmeier1995,feldmeier97}, at least part of the flares could be due to drops of the wind speed and consequent increases of the accretion radius. These flares would not be necessarily associated to column density enhancements.

In the cases of IGR\,J16479$-$4514 and IGR\,J18450$-$0435, we could not investigate in detail the HR variations due to the complication of the egress from the eclipse and the low statistics of data, respectively. Previous publications in the literature have shown that these sources can display extreme variability \citep[see, e.g., BZ17, ][and references therein]{Sidoli2006, zurita09, Sidoli2017}, and thus additional observations with \xmm\ in the future might help us catching bright flares and achieving a better understanding of the physical conditions in their accretion environments.

Two sources in our sample were not detected during the corresponding \xmm\ observation. In the case of IGR\,J11215$-$5952, we determined a 90\% c.l. upper limit of
6$\times$10$^{-15}$~erg~cm$^{-2}$~s$^{-1}$ on its 1--10\,keV unabsorbed X-ray emission, corresponding to a luminosity of 4$\times$10$^{31}$~erg~s$^{-1}$
\citep[assuming a distance of 7~kpc; see][and references therein]{Sidoli2017}. At the best of our knowledge, this is the lowest flux ever recorded from this source \citep[see also][for recent non detections]{Sidoli2020}. This low luminosity would not
be surprising, as IGR\,J11215$-$5952 is known to have a long and eccentric orbit which would naturally cause the mass accretion rate to drop dramatically when the neutron staris far away from
periastron. As the orbital period of the source is $\sim$165~days and the last accurately observed outburst occurred on 2016 February 14, we conclude that the \xmm\ observation took place about  60~days after the closer expected outburst, likely in a region of the orbit where the mass accretion rate is too low to give rise to a detectable X-ray emission. Although we only have one deep
upper limit so far, the availability of further high-sensitivity observations with \xmm\ along the orbit of IGR\,J11215$-$5952 might help us making a full comparison with the case of neutron starBe X-ray
binaries which are endowed with similar high eccentricity and elongated orbits but which are often detected in X-rays (including pulsations) also during time intervals away from periastron
\citep[see][and references therein]{D14}. Due to its peculiar orbital configuration and the uniqueness of its periodic outbursts among the SFXT sources, it was already claimed in previous papers
that IGR\,J11215$-$5952 could be the missing link between SFXTs and the longer known class of Be X-ray binaries \citep{liu11}.

The estimated upper limit on the X-ray emission from IGR\,J18410$-$0535 measured during the \xmm\ observation 0844100101 provided, at the best of our knowledge, again the lowest luminosity value
for this object \citep[see][and references therein]{sidoli08,bozzo11}. As our upper limit is a factor of $\sim$10 deeper than previously reported values, this result increases the dynamic range
displayed by IGR\,J18410$-$0535 up to $\sim$5$\times$10$^4$. This is still well within the dynamic ranges displayed by the SFXTs, which record of $\sim$10$^{6}$ was achieved by IGR\,J17544$-$2619
\citep{romano15b}.

The summary above shows that the monitoring program of the SFXTs that we are pursuing with \xmm\ is already providing intriguing and useful results, according to the expectations. Further \xmm\ observations of a similar duration as those reported here are currently on-going for a number of other confirmed SFXT sources. When a large amount of flares per source is available ($\gtrsim$25), it will be possible to
perform more accurate statistical analyses on the properties of the fast spectral variability during these events across the entire SFXT class, as suggested and initiated by BZ17. This will be a powerful instrument to improve significantly our understanding of the mechanisms triggering the
peculiar X-ray variability of SFXTs, possibly going quantitatively beyond the simplistic assumption that most of the variability is related to stellar wind clumps and providing reliable estimates of the physical properties of these structures in both cases when they are either being accreted by the neutron star or simply passing along our line of sight to the compact object.

\begin{acknowledgements}
    We thank the anonymous referee for their useful comments which helped us improving the paper. This research made use of \texttt{Astropy},\footnote{\url{http://www.astropy.org}} a community-developed core Python package for Astronomy \citep{astropy2018}. For plotting, we exploited the \texttt{Matplotlib} python package \citep{Hunter:2007}. To consistently perform the analysis, we developed a python wrapper to particular SAS functions, which is publicly available: \href{https://gitlab.astro.unige.ch/ferrigno/pysas}{pyxmmsas}.
\end{acknowledgements}

\bibliographystyle{aa}
\bibliography{reference}

\end{document}